\documentclass[]{spie}  

\pdfoutput=1

\usepackage{graphicx}
 
\usepackage{lmodern} 
\usepackage[T1]{fontenc} 
\usepackage{textcomp}
\graphicspath{{pic/}}

\usepackage{amsmath,amsfonts,amssymb}
\usepackage[colorlinks=true, allcolors=blue]{hyperref}

\title{Minority carrier diffusion lengths and mobilities in low-doped
   n-InGaAs for focal plane array applications}

\author[]{Alexandre W. Walker}
\author[]{Mike W. Denhoff}
\affil[]{National Research Council of Canada, 1200 Montreal Road,
            M-50, K1A 0R6, Ottawa, Ontario, Canada.
    May 2017.\\ 
   {\small Published in: Infrared Technology and Applications XLIII,
    edited by Bjørn F. Andresen,
    Gabor F. Fulop, Charles M. Hanson,
    John Lester Miller, Paul R. Norton, Proc. of SPIE Vol. 10177, 101772D.}}

\authorinfo{Further author information: (Send correspondence to A. W. Walker)
            \\A. W. Walker: E-mail: alexandre.walker@nrc-cnrc.gc.ca
             \\M. W. Denhoff: E-mail: mwdenhoff@gmail.com}

\begin{document} 
\maketitle

\thispagestyle{plain}
\begin{abstract}
The hole diffusion length in n-InGaAs is extracted for two samples of different doping
concentrations using a set of long and thin diffused junction diodes separated by various distances on the order of the
diffusion length. The methodology is described, including the ensuing analysis which yields diffusion
lengths between 70\,-\,85\,\textmu m at room temperature for doping concentrations
 in the range of $5\,-\,9\times10^{15}\,\mathrm{cm}^{-3}$. The analysis also provides insight into the minority carrier mobility which is a parameter not
commonly reported in the literature.
 Hole mobilities on the order of $500\,-\,750$ cm$^2$/V$\cdot$s are reported for
the aforementioned doping range,
 which are comparable albeit longer than the majority hole mobility
for the same doping magnitude in p-InGaAs.
 A radiative recombination coefficient of $(0.5\pm 0.2)\times10^{-10}$\,cm$^{-3}$s$^{-1}$ is
 also extracted from the ensuing analysis for an InGaAs thickness of 2.7\,\textmu m.
 Preliminary evidence is also given for both heavy and light hole diffusion.
 The dark current of InP/InGaAs \textit{p-i-n} photodetectors
with 25 and 15\,\textmu m pitches are then calibrated to device simulations
 and correlated to the extracted
diffusion lengths and doping concentrations.
 An effective Shockley-Read-Hall lifetime of between 90-200\,\textmu s provides the best fit to the dark current of these structures.

\end{abstract}


\section{INTRODUCTION}
\label{sec:intro}  

    Minority carrier devices such as photodetectors, solar cells, light-emitting-diodes and bipolar
junction transistors rely on sufficiently long minority carrier diffusion lengths for high performance.
However, these values are not commonly reported in the literature due to the lack of simple methods to
extract these values. Common methods include cathodoluminescence
 measurements\cite{Zaram, Schultes, Boulou, Gustafsson}, beam
induced current via a controlled light source\cite{Sharma},
 zero time of flight measurements\cite{Lovejoy} or
surface photovoltage measurements\cite{Scroder,Kronik}, all of which are typically conducted on double-
heterostructures or cross-sections of a device, and therefore not on the final devices. These methods
require either an electron beam such as from a scanning electron microscope, or a calibrated light
source for constant photogeneration such as a laser (with appropriate filters for low injection). In any
case, a suitably designed contact mask is required, or a calibrated photodetector coupled to a simple
contact mask. All of these introduce some level of complexity as well as uncertainties in the ensuing
analysis. Another method involves modeling the responsivity of a device as a function of wavelength
compared to experiment\cite{Walkera},
 but this requires accurate datasets of the optical properties, and only
provides a lower bound on the diffusion length if the diffusion length is sufficiently long compared to the
active region thickness. These aforementioned methods each have their own advantages, but more
importantly their own inherent limitations and complexities; as a result, limited reports exist of minority
carrier diffusion lengths in the literature for III-V semiconductors, as well as their dependencies, such as doping. Most notably is the lack of minority carrier mobilities reported, which require independent
measurements of minority carrier diffusion lengths and minority carrier lifetimes (often via time-
resolved photoluminescence which requires sophisticated sub-nanosecond detectors). Furthermore,
both of these independent measurements must be made at the same injection level due to the injection
level dependence of minority carrier lifetimes\cite{Walkerb, Ahrenkiela}. As a result, minority carrier mobilities are often
assumed to be equivalent to the majority carrier mobility in the oppositely doped material (i.e. holes in
n-InGaAs have the same mobilities as holes in p-InGaAs), which has been shown to be a poor
approximation in the limited cases reported such as in GaInP\cite{Schultes}, although that may not always be the case,
for example, in GaAs\cite{Lovejoy}.

    In this paper, a simple method of extracting the diffusion length of minority carriers and their
respective mobilities is described which uses solely electrical measurements of a set of long and thin diffused junction diodes
(``line diodes'', see Figure 1)\cite{Kopanski}. These line diodes can be easily integrated into test structures to monitor process
controls in both epitaxy and fabrication. The methodology and corresponding theory is described in
section 2. Section 3 then reports the extracted diffusion lengths of holes in n-InGaAs as a function of the
injected carrier concentration for two different doping concentrations. The extracted minority carrier
mobilities are also reported for these samples. Section 4 then correlates the dark current of 100 pixel
test arrays to the extracted diffusion lengths and doping concentrations. Finally, section 5 gives the conclusions of the study.

   \begin{figure} [t]
   \begin{center}
     (a) \underline{Top view}
   \includegraphics[]{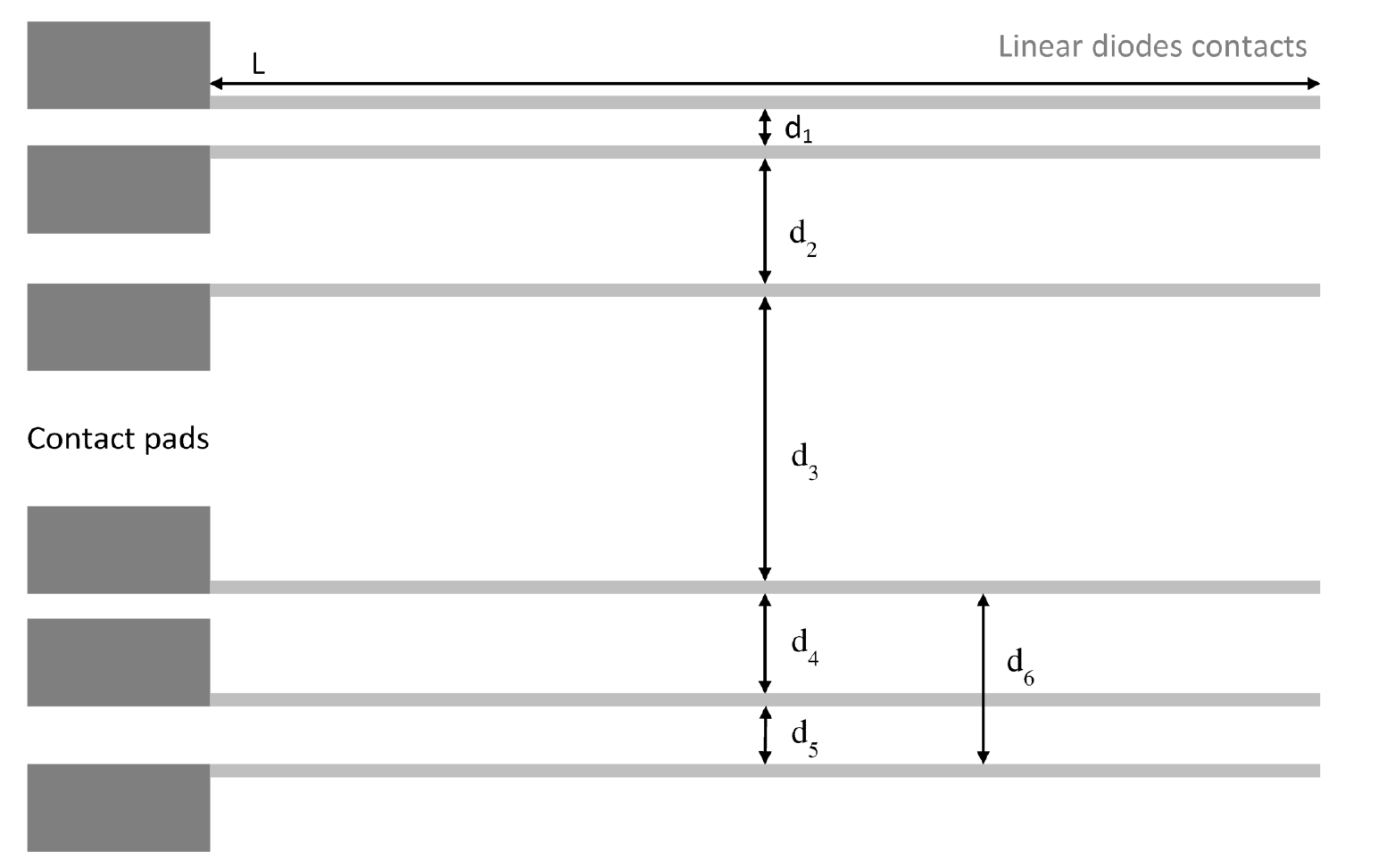}\\
    (b)  \underline{Cross-sectional view of double-heterostructure with diffused p-region}\\[1eX] 
	\includegraphics[width=0.9\textwidth]{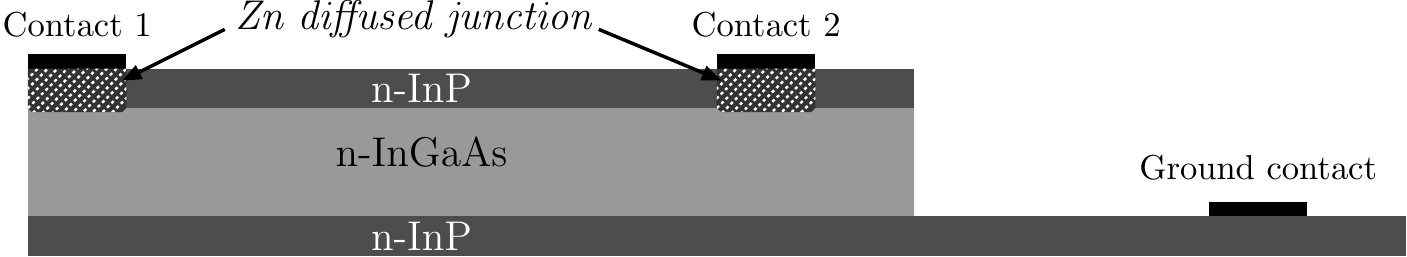}
   \end{center}
   \caption[example]
   { \label{fig:fig-1}
 a) Schematic of a set of long and thin diffused junction diodes of length L with various
 inter-diode spacings (d1, d2, ..., d6) to extract minority
carrier diffusion lengths solely using electrical measurements.
 The diode length L must be significantly larger than the largest
inter-diode spacing (L $\gg$ max(d)). b) Cross-section of two 
line diodes and the corresponding ground, revealing the InP/InGaAs/InP epitaxial stack.}
   \end{figure}

\section{Methodology}
\label{sec:meth}

   \begin{figure} [t]
   \begin{center}
   \begin{tabular}{cc} 
   \includegraphics[width=0.48\textwidth]{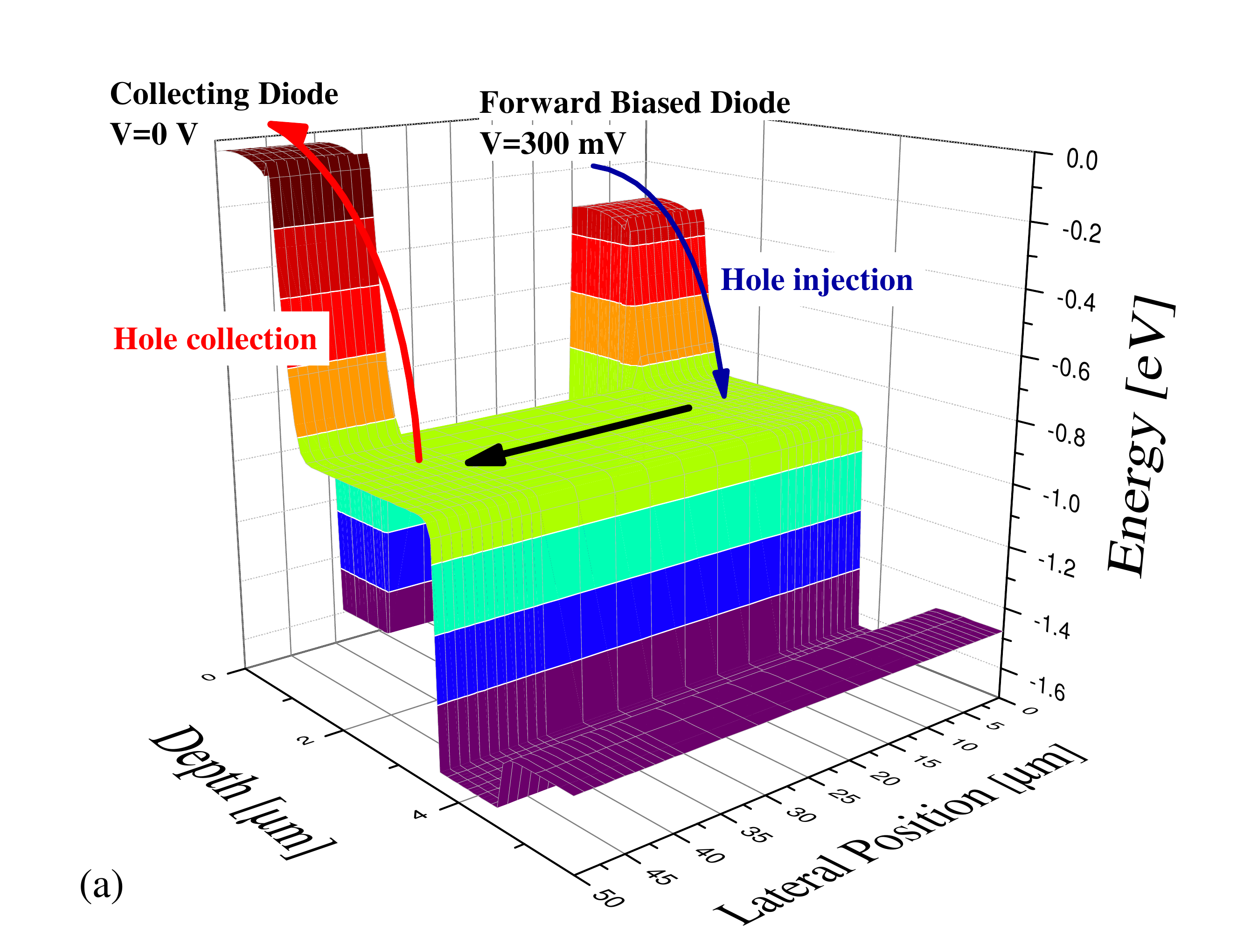} &
   \includegraphics[width=0.48\textwidth]{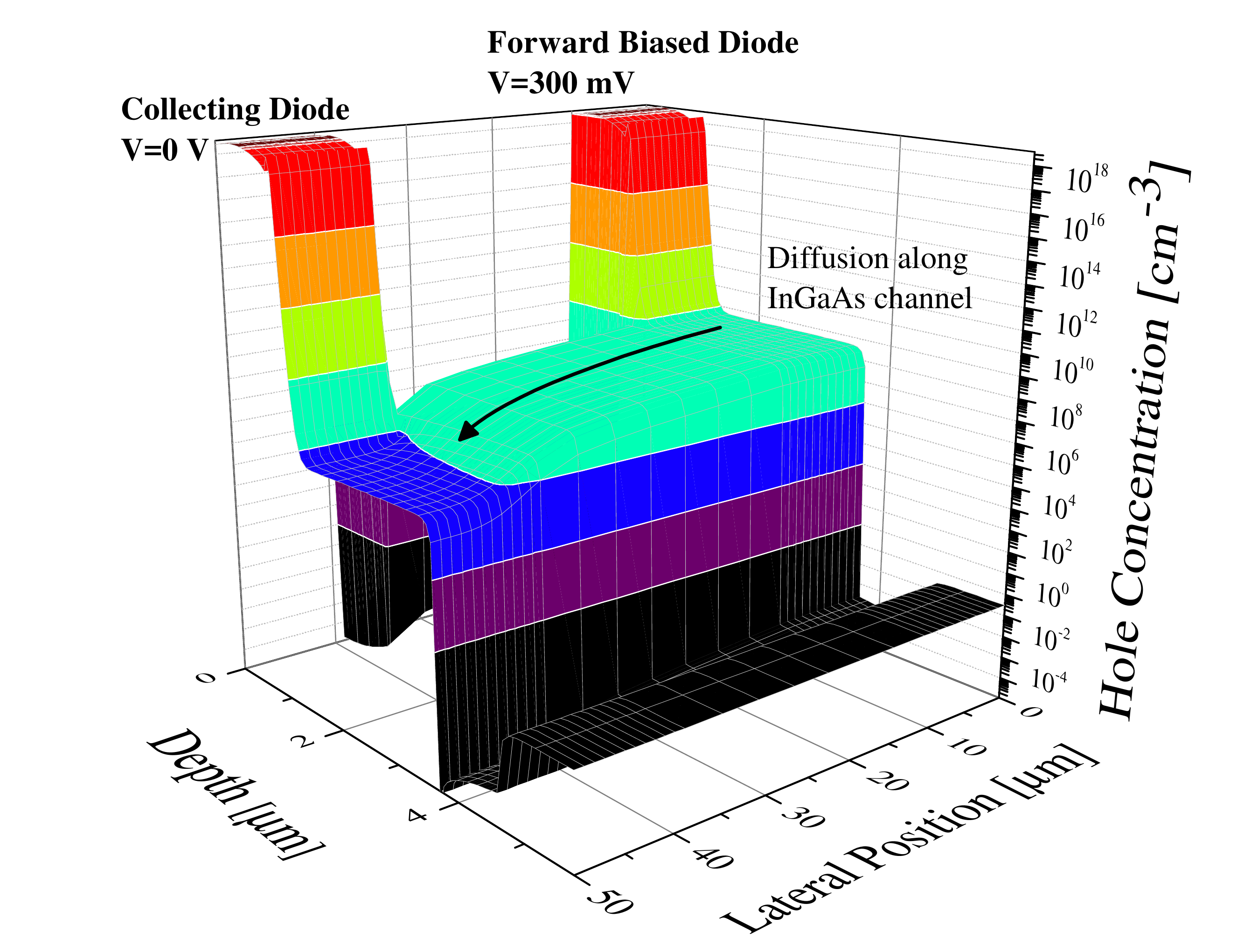}
   \end{tabular}
   \end{center}
   \caption[example]
   { \label{fig:fig-2}
 Simulated a) valence band diagram and b) hole concentration
 of the InGaAs two-diode configuration at a 300\,mV
forward bias (for the injection diode). This demonstrates that holes
 injected into the InGaAs layer are confined along this
channel until reaching the collecting diode.}
   \end{figure}

    Figure~\ref{fig:fig-1}a illustrates a schematic of how these line
 diodes could be setup, where inter-diode
distances ideally correspond to 0.3\,-\,5 times the diffusion length. The number of diodes should be on the
order of 4-6 for a reasonable analytical fit.
 Forward biasing one line diode injects minority carriers into
the InGaAs layer beneath the first diode, leading to diffusion of carriers in the InGaAs channel (see
Figure~\ref{fig:fig-1}b for a cross-section of the device).
 The carriers which diffuse to a second line diode can be
collected by applying a zero or slightly reverse bias to that diode. If the length of the line diodes L is
sufficiently long compared to the maximum inter-diode distance d$_i$, then a one-dimensional
approximation (along $x$) can be made to solving the hole continuity
 equation in the case of a fixed hole injection level:
\begin{equation}
\label{eq:one}
\frac{\partial p}{\partial t} = G-R - \frac{1}{q}\frac{\partial J}{\partial x} = 0 ,
\end{equation}
where $\rho$ is the hole concentration, $G$ and $R$ are the generation and
 recombination rates respectively, $J$ is
the current density and $q$ the electronic charge. Assuming steady state and no generation, i.e. no light is
illuminated on the device ($G = 0$), equation~(\ref{eq:one}) simplifies to
\begin{equation}
\label{eq:two}
\frac{1}{q}\frac{\partial J}{\partial x} = R = \frac{p-p_0}{\tau},
\end{equation}
where $\tau$ is the minority carrier lifetime, and the recombination
 rate is approximated for low injection,
i.e. $R = \frac{p-p_0}{\tau}$.\cite{Sze}
  The  current  density  is  given  by  drift  and  diffusion  contributions  as
\begin{equation}
\label{eq:three}
J = qp\mu \frac{\partial E}{\partial x}+qD\frac{\partial^2 p}{\partial x^2},
\end{equation}
where $\mu$ is the hole mobility, $E$ is the electric field, and $D$ is the
 hole diffusion coefficient. Substituting equation (3) into equation (2) leads
 to the one-dimensional current-continuity equation for holes
\begin{equation}
\label{eq:four}
\frac{p-p_0}{\tau}+p\mu \frac{\partial E}{\partial x}
   +\mu E\frac{\partial p}{\partial x}+D\frac{\partial^2 p}{\partial x^2}=0.
\end{equation}

    In order to illustrate this one-dimensional diffusion of holes in the structure,
 the valence band of two line diodes is
simulated in 2D using the numerical device solver
 Atlas by Silvaco (version 5.2.2.1.R, Santa Clara, CA,
USA, 2016) using the built-in properties for InP and InGaAs
 lattice matched to InP, and the device geometry
corresponding to Figure~\ref{fig:fig-1}b. Figure~\ref{fig:fig-2}a
 illustrates the valence band diagram for the injecting diode
operating at a forward bias of 300\,mV.
Minority carriers are injected into the InGaAs by the injecting diode and are subsequently confined in the InGaAs channel as they diffuse to the collecting diode.
 The hole concentration under these particular operating conditions is shown in
 Figure~\ref{fig:fig-2}b, and clearly illustrates a
diffusion process along the InGaAs channel.
 Since the applied bias is dropped solely across the injecting
diode (as seen in Figure~\ref{fig:fig-2}a by the lowering of the
 valence band at the injecting diode by 300\,meV
compared to the collecting diode), the current-continuity equation along this InGaAs channel can be
simplified to
\begin{equation}
\label{eq:five}
\frac{p-p_0}{\tau}+D\frac{\partial^2 p}{\partial x^2}=0.
\end{equation}
This can also be expressed as
\begin{equation}
\label{eq:six}
\frac{\partial^2 p}{\partial x^2} = \frac{p-p_0}{\tau D}.
\end{equation}
The general solution to this equation is
\begin{equation}
\label{eq:seven}
p(x) = C_1\sinh(x/L_D)+C_2\cosh(x/L_D)+p_0,
\end{equation}
where $L_D = \sqrt{\tau D}$  is the diffusion length of the minority carriers,
 and $C_1$ and $C_2$ are constants based on
boundary conditions of the problem. At $x = 0$, the injecting diode has a
 hole concentration of $p(x\!=\!0)= p_{inj} = n_i^2/ N_D\exp(qV/k_BT)$
 based on a low injection (Boltzmann) approximation. According to Figure~\ref{fig:fig-2}b, the hole population across the thickness of the InGaAs layer (along $x = 0$) is quite homogeneous,
 thus validating this assumption. At
the second diode $(x\!=\!W)$, the concentration is $p(x\!=\!W) = p_0 = n_i^2/ N_D$,
 because the second diode collects all
carriers that reach its space charge region, and thus the equilibrium
 carrier concentration remains. This assumption can also be verified by
 investigating Figure~\ref{fig:fig-2}b which shows a hole concentration in the
InGaAs directly below the collecting diode 
of $p(x = W) = 7.9\times 10^{7}\mathrm{cm}^{-3}\simeq  p_0$. The first boundary
condition leads straightforwardly to $C_1 = p_{inj} - p_0$,
 whereas the second boundary condition leads to
$C_2 = -(p_{inj} - p_0)/\tanh(W/L_D)$.
 Therefore, the current density at the collected diode $(x=W)$ can be
expressed as a function of inter-diode separation $W$ as
\begin{equation}
\label{eq:eight}
J_p(W) = -qD\left. \frac{\partial p}{\partial x}\right|_{x=W}
   = -\frac{qD(p_{inj}-p_0)}{L_D}
  \left[\sinh\left(\frac{W}{L_D}\right)-\frac{\cosh(W/L_D)}{\tanh(W/L_D)}\right]
   = \frac{qD(p_{inj}-p_0)}{L_D\sinh(W/L_D)}.
\end{equation}
This derivation agrees with that given in\cite{Sze} for the decay of excess
 carriers over distance with the special second boundary condition of
 carrier extraction at $x = W$.
 Measuring the current at the second
diode for various inter-diode distances therefore reveals both the diffusion length and the diffusion
coefficient  based on a best-fit of equation~(\ref{eq:eight})
 to the measured data, assuming the injected hole
concentration can be modeled using Boltzmann statistics
 (which is valid under low injection). Since the
diffusion coefficient is directly linked to the minority carrier
 mobility according to the Einstein relation,
one can then extract this parameter in the ensuing analysis and compare
 it to the hole mobility in $p$-type
material for the equivalent doping.

\section{Minority Carrier Properties}

\begin{figure} [t]
   \begin{center}
   \includegraphics[width=0.55\textwidth]{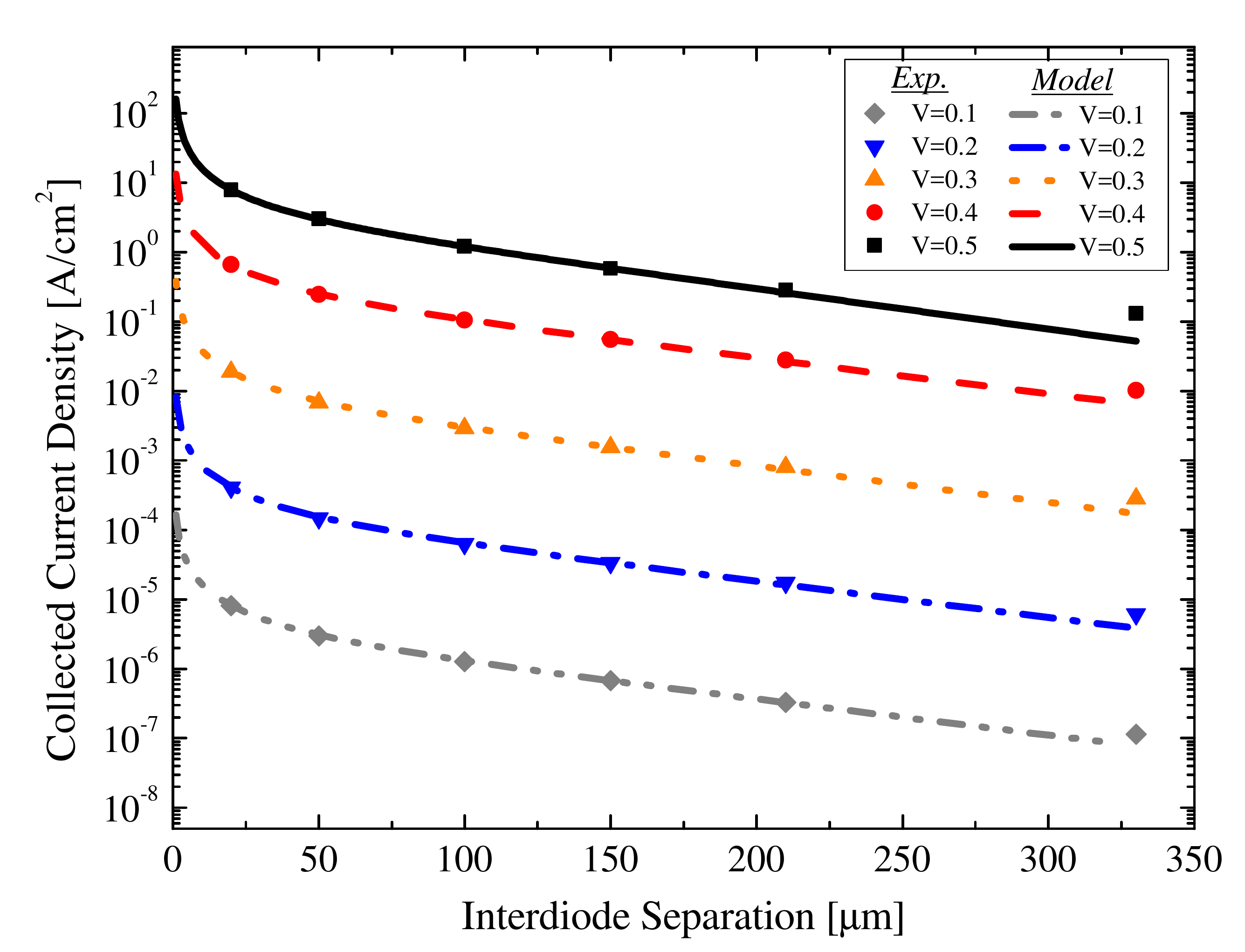}
   \end{center}
   \caption[example]
   { \label{fig:fig-3}
Measured current collected at the secondary diode as a function of
 interdiode separation, along with best-fit
calculation to equation~(\ref{eq:eight}), for various voltages.}
   \end{figure}

    Figure~\ref{fig:fig-3} illustrates the measured current density collected at the secondary diode for increasing forward biases
at the first diode as a function of the inter-diode separation. A constant 0\,V bias is maintained at the
secondary diode, and a common 0\,V bias applied to the ground contact. Note that Figure~\ref{fig:fig-3} shows
current densities and not absolute currents.
 The best-fit to equation~(\ref{eq:eight}) for each data set also shows a
very reasonable fit, as demonstrated by an adjusted R-squared$\,>\,$0.999 using Matlab R2016b's curve
fitting toolbox with a nonlinear least squares fitting routine based on the Levenberg-Marquardt
algorithm, which also generates insight into the uncertainties from the fitting procedure. However, for
the longest inter-diode distance explored of 330\,\textmu m, the fit clearly deviates from the measurement,
which one may conclude is due to the 1D approximation failing (i.e. $L\sim d$) which leads to non-1D
diffusion; however, one would then expect that the measured collected current would be less than the
1D model which is not the case. This leads to the hypothesis that this higher current collected at the largest interdiode separation is due to
light holes having significantly longer diffusion lengths than heavy holes, whereas the collected current for the
smaller interdiode distances is dominated by heavy hole diffusion. This is supported by the $\sim8$x
lower light hole effective mass compared to heavy
 holes\cite{Hermann,Levinshtein}, which has two repercussions: 1) the heavy hole
band dominates the valence band density of states, resulting in heavy holes dictating the current for short interdiode separations, and 2) light hole diffusion lengths are longer than heavy hole diffusion lenghts, thus resulting in light holes dominating the largest interdiode separations. This has a significant impact
on the extracted diffusion length, since the data should be fit
 by two variations of equation~(\ref{eq:eight}): one for
heavy holes and one for light holes. The reported diffusion lengths and mobilities are therefore
representative of heavy holes, and should be interpreted as upper limits. This hypothesis merits further study.

  \subsection{Diffusion Lengths}

\begin{figure} [t]
   \begin{center}
   \begin{tabular}{cc} 
 \hspace{-1em}  \includegraphics[width=0.51\textwidth]{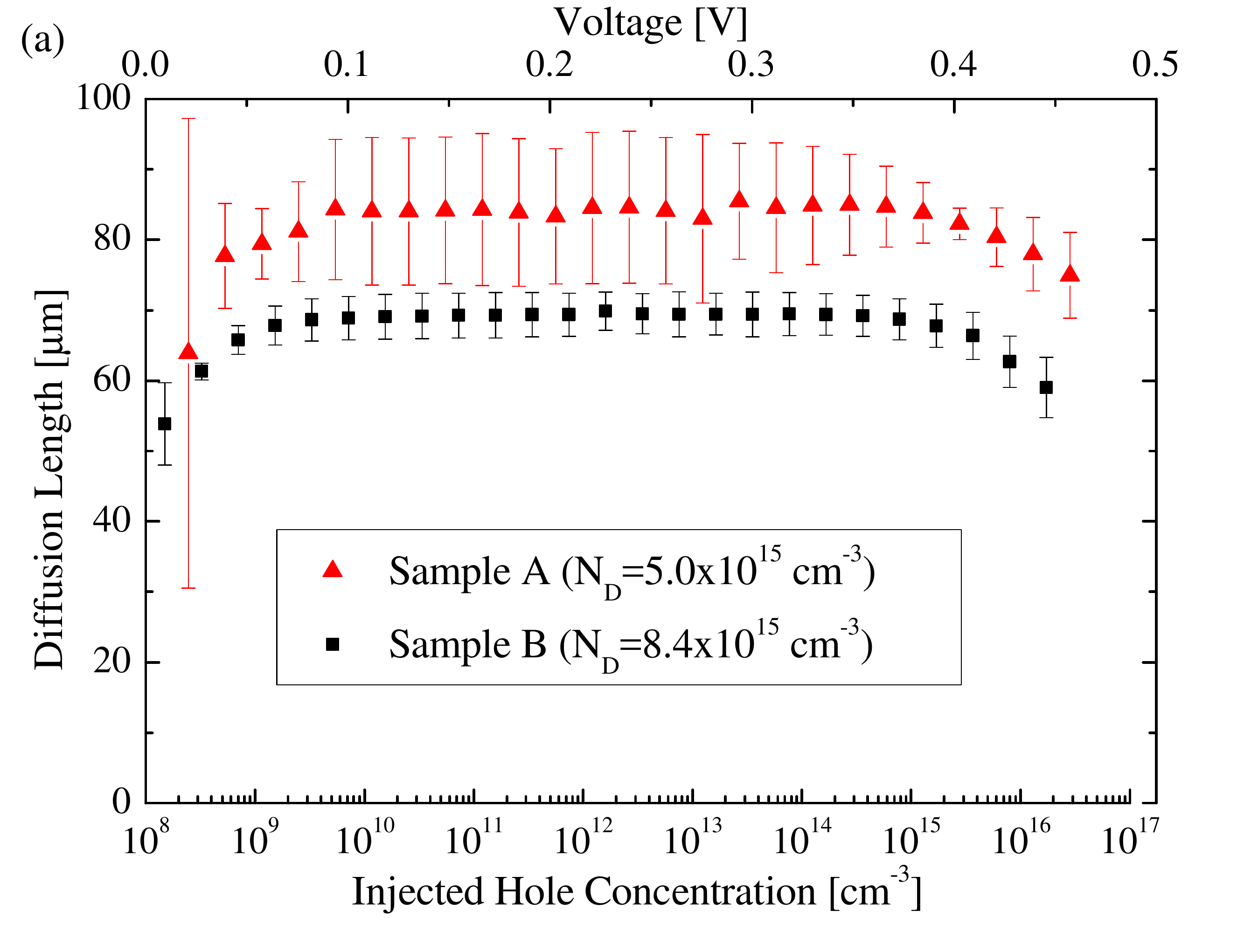} &
 \hspace{-2em}  \includegraphics[width=0.51\textwidth]{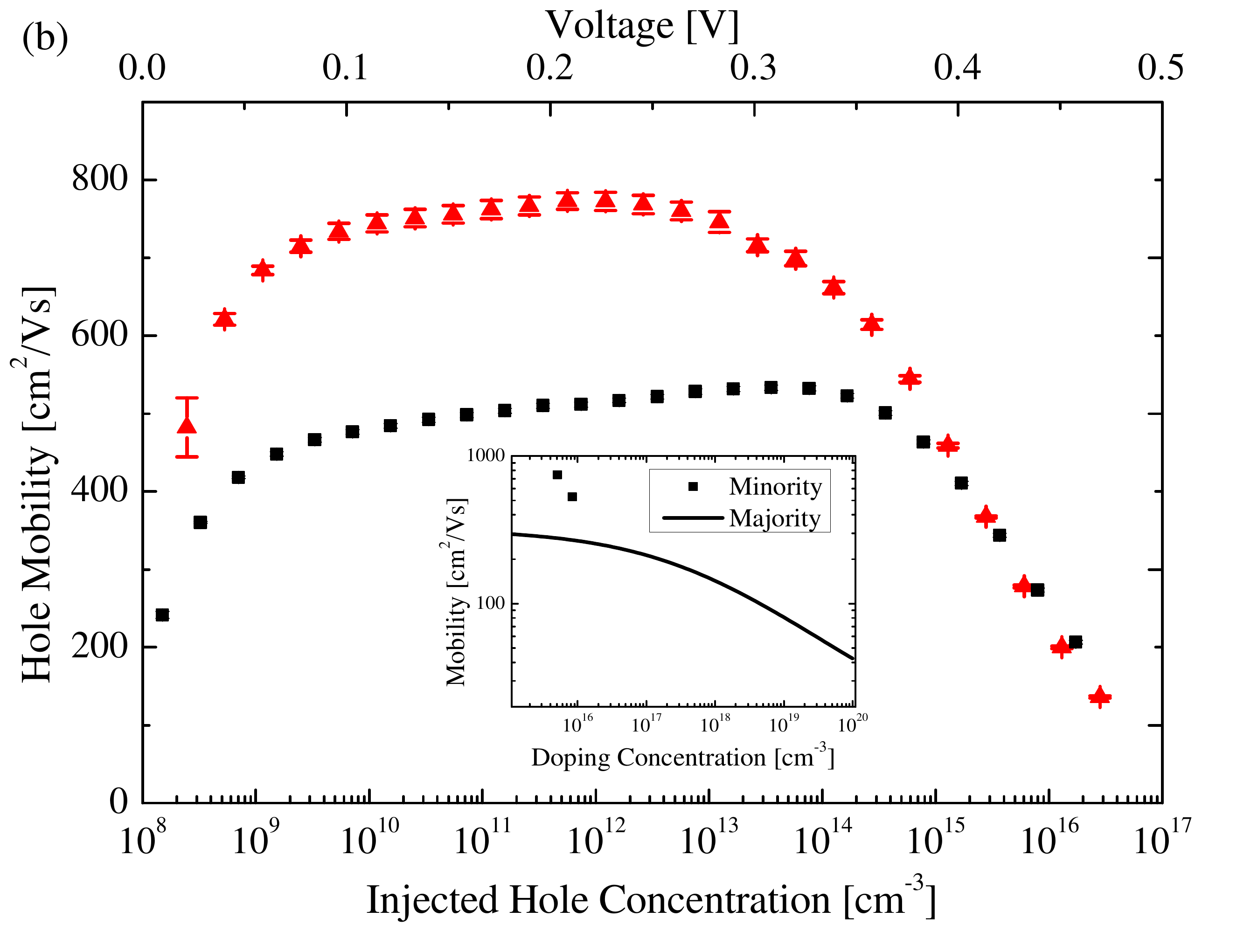}
   \end{tabular}
   \end{center}
   \caption[example]
   { \label{fig:fig-4}
 Extracted minority carrier a) diffusion lengths, and b) mobilities as a
 function of injected carrier concentration for
two samples of different carrier concentrations (injecting voltage on top axis).
 Inset shows the hole mobility as extracted from this study for
 minorities for a voltage of 0.3\,V, and the majority carrier mobility of
 p-InGaAs both as a function of doping\cite{Sotoodeh}.}
   \end{figure}

    The best-fit diffusion length as a function of injected carrier
 concentration are illustrated in Figure~\ref{fig:fig-4}a
 for two samples of different doping levels:
 Structure A with $5.1\times10^{15}\,\mathrm{cm}^{-3}$, and Structure B with
$8.4\times10^{15}\,\mathrm{cm}^{-3}$ as extracted from capacitance - voltage
 measurements on 200\,\textmu m diameter diffused junctions; note the voltage is illustrated on the upper axis.
 The very low injection conditions ($V<0.05$ V) shows more scatter
 due to the very low currents of $1\,-\,10\,$ pA and due to hysteresis in the measurement.
 This therefore gives rise to a larger uncertainty than in in the higher voltage
measurements which have exponentially higher currents and are less impacted by hysteresis.
 In general, there is a constant diffusion lengths as a function of injection
 of 70\,\textmu m and 85\,\textmu m for both samples up to an injected hole
concentration of close to $1\times10^{15}$\,cm$^{-3}$.
 After this voltage, there is a trend of decreasing diffusion length,
which originates from the impact of the high injection
 regime on the radiative lifetime; this also
invalidates equation~(\ref{eq:eight}).
 These diffusion lengths are shorter compared to another reported finding on
hole diffusion lengths in n-InGaAs,\cite{Gallant} most likely due to the lower
 doping of $10^{15}$\,cm$^{-3}$ which would lead to
a lifetime $5\,-\,8$ times longer than in this study.
 However, the adopted diffusion coefficient
and radiative recombination coefficient in a study by Wichman \textit{et al.}\cite{Wichman} 
leads to a diffusion length of 30\,\textmu m, which is nearly half that
 reported here yet for the same doping range. This highlights the need
for more reports of diffusion lengths in the literature.

   \subsection{Mobilities}

    The best-fit also gives insight into the diffusion coefficient $D$,
 which in turn gives the minority carrier
mobility according to the Einstein relation $D = \mu k_BT/q$.
 This data is shown in Figure~\ref{fig:fig-4}b as a function of
injected hole concentration. In general, there is a constant mobility
 as a function of injection of 500 and
750 cm$^2$/V$\cdot$s for samples A and B respectively.
 However, the results show a reasonably sharp increase in
mobility in the lowest injection range of $10^{8}\,-\,10^9$\,cm$^{-3}$,
 and finally terminated by an even sharper
decrease in mobility in the high injection range. The initial sharp increase in mobility is due to the hystersis of the measurement, which results in a non-zero voltage offset that impacts the fitting procedure. The extracted minority carrier mobility in the lowest
injection level, with the largest uncertainties, should therefore not be trusted unless hysterisis is properly removed from the measurement, by starting the current - voltage measurement at 0 V for example.
 However, above a voltage of $V=0.05$\,V, the mobility becomes
trustworthy, at least until the high injection regime begins. At this point, the minority carrier
 mobility decreases
considerably, which is due to two factors: first, the current collected becomes limited due to series
resistance in the structure, and second, the proposed analytical model fails to account for majority
carrier diffusion which begins to contribute to the collected current density. Nevertheless, the resulting
minority carrier mobility (in the middle injection range) can be compared to the majority carrier mobility
as a function of doping,\cite{Sotoodeh} as shown in the inset of Figure~\ref{fig:fig-4}b.
 The relative error between the two sets
of mobilities ranges between $95\,-\,170$\% with respect to the Sotoodeh data. This emphasizes the need for
more accurate minority carrier mobilities reported in the literature. The presented methodology is a
very useful and simple method that allows for straightforward reporting of such values. Note that the doping concentration in the InGaAs
must be known accurately to estimate the injection level and thus the mobilities. This adds an extra source of uncertainty in the extracted mobility values.

     \subsection{Radiative Recombination Coefficient}
    Based on the extracted minority carrier diffusion length and mobility, the lifetime of the carrier can
then be worked out which gives insight into the bulk InGaAs radiative recombination coefficient. The
diffusion length equation is given by $L_D = \sqrt{D\tau}$ ,
 where the minority carrier recombination lifetime is
dictated by radiative, Shockley-Read-Hall (SRH) and Auger processes as
$\tau  = (1/\tau_{RAD} + 1/\tau_{SRH} + 1/\tau_{AUGER})^{-1}$. In the low
injection level, Auger can be assumed to be negligible, and if SRH is not important (this is justified in section 4), such that one can then isolate the radiative
recombination coefficient ($B_{RAD}$) from $\tau_{RAD} = 1/B_{RAD}N_D$.
 The result is $(0.5\pm 0.2)\times10^{-10}$\,cm$^{-3}$s$^{-1}$ for both samples,
which is in better agreement with the value of $0.4\times10^{-10}$\,cm$^{-3}$s$^{-1}$
 reported by Wintner \textit{et al.}\cite{Wintner} than the
typical values of $0.8\,-\,1\times10^{-10}$\,cm$^{-3}$s$^{-1}$ adopted by
 Wichman \textit{et al.}\cite{Wichman} or reported experimentally\cite{Zielinski,Ahrenkielb}. However, the thickness of the InGaAs layer has a critical influence on the
amount of photon recycling which directly impacts the radiative
 recombination coefficient\cite{Walkera,Dagan,Lumb}.
 It is thus reasonable to have a variety of reported radiative
recombination coefficients as this depends primarily on thickness, which is a parameter that is
unfortunately not reported in the experimental study of Ahrenkiel \textit{et al.}\cite{Ahrenkielb}

   \section{Diffusion Length Correlation to Dark Current}

    The influence of doping and the corresponding diffusion lengths on the dark current of 100 pixel photodetector test arrays is
investigated in this section using the experimental data as well as simulations again using Atlas by Silvaco for both structures of interest. The simulation
models pixel pitches of 25 and 15\,\textmu m in cylindrical coordinates based on symmetry elements
corresponding to a radius of 12.5\,\textmu m and a 7.5\,\textmu m respectively.
 A radiative recombination coefficient of
$0.7\times10^{-10}$\,cm$^{-3}$ is assumed based on the previous calculations, and
 since this provides the best agreement to
the forward bias diffusion current.
 The Zn apertures are 6\,\textmu m and 2\,\textmu m in radii respectively. Each pixel
of the test arrays consists of the InP/InGaAs/InP double-heterostructure, where the \textit{p-i-n} junction is
realized using a diffusion of Zn at the top InP layer (see Figure~\ref{fig:fig-1}b).
 The Zn diffusion profile is simulated
using the Silvaco module Athena (v. 5.22.1.R) using the same diffusion time and temperature as used
experimentally. The Zn diffusion parameters in InP were chosen to match the experimental Zn diffusion
profile. The Zn penetrates into the InGaAs by $\sim$100\,nm.

\begin{figure} [t]
   \begin{center}
   \begin{tabular}{cc} 
 \hspace{-1em}  \includegraphics[width=0.52\textwidth]{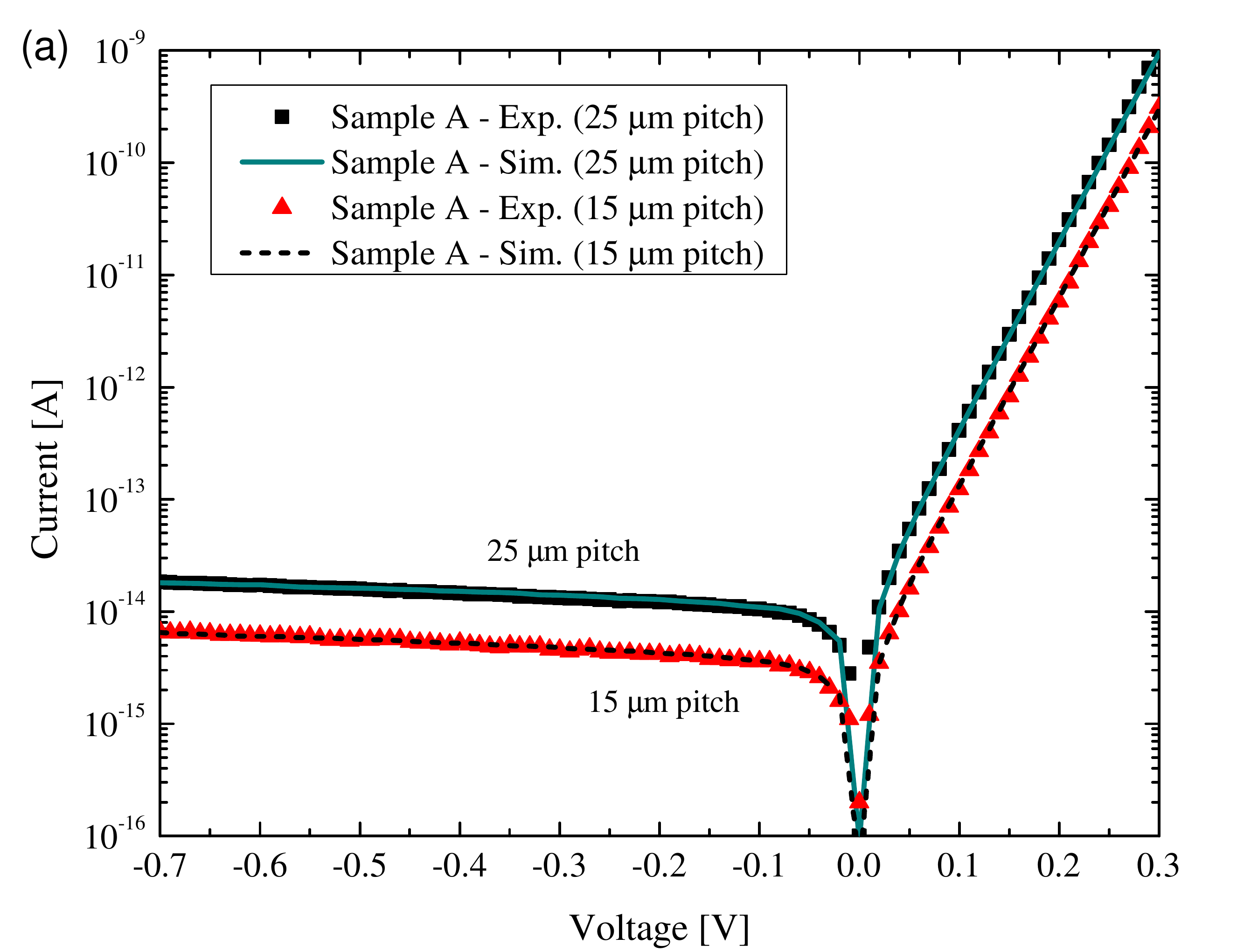} &
 \hspace{-2em}  \includegraphics[width=0.52\textwidth]{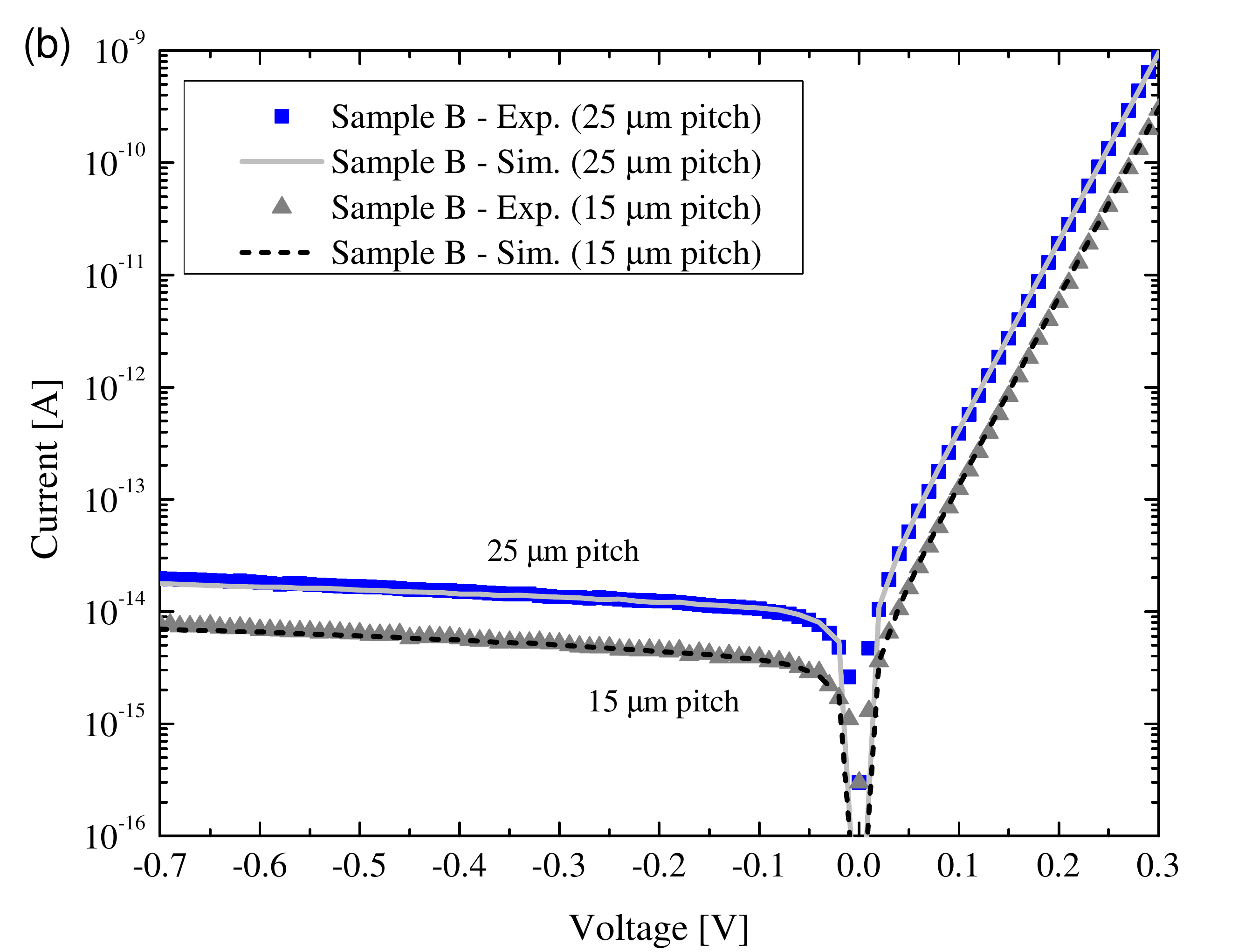}
   \end{tabular}
   \end{center}
 \vspace{-2ex}  \caption[example]
   { \label{fig:fig-5}
Model calibration for a) sample A and b) sample B, each with 25 and 15\,\textmu m pitches.
}
   \end{figure}

    The measured dark current densities of an average pixel based on 100 pixel test arrays with 25 and
15\,\textmu m pitches are shown in Figure~\ref{fig:fig-5}a and b
 respectively for both samples, along with the model
calibrated to the measurements. The measurements reveal that both structures have very comparable
dark currents, even though the diffusion lengths in structure B are $25\,-\,30$\% longer than in structure A,
and that the doping in structure B is nearly 50\% higher than in structure A. The increased doping leads
to a reduction in depletion region width of about 100\,nm (from 550 to 450\,nm total), which reduces the
overall contribution of SRH in the structure (since SRH is proportional to the depletion region volume).
However, this reduction in depletion results in an increase in diffusion in the neutral InGaAs layer. The net change is close to
zero, thus balancing junction recombination with diffusion.

    According to the modeling, a hole SRH lifetime of 200\,\textmu s in the InGaAs material provides a
reasonably good fit for both of our structures with 25\,\textmu m pitch,
 which is longer than the results from an experimental study on the various
 components of recombination in InGaAs.\cite{Ahrenkiela} Note that
this modeling study does not explicitly account for interface recombination at the InP/InGaAs nor at the
InP surface. For the smaller pitch devices, a shorter lifetime of 90\,\textmu s yielded the best fit. This shorter
lifetime implies that perimeter effects are more important for smaller pitch devices, as expected based
on previous studies\cite{Wichman,MacDougal,Yuan}.
 For both of our structures, approximately
1/2 of the total dark current in reverse bias corresponds to generation-recombination at the junction via
SRH recombination, whereas the remaining 1/2 is due to diffusion in the neutral InGaAs region. This
suggests there is room for improvement in minimizing the dark current further by inhibiting SRH
processes at the junction. These relative contributions are in reasonable agreement with a recent similar
modeling studies corresponding to comparable \textit{p-i-n} photodetector arrays\cite{Wichman}.
 At a reverse
bias of 0.5\,V at room temperature, the dark current densities are 2.5 nA/cm$^2$
 for both 25 and 15\,\textmu m
pitches in structure A, whereas structure B's values
 are very similar at 2.6 and 2.7 nA/cm$^2$
for 25 and 15\,\textmu m pitches respectively.
 These values are also quite comparable to the literature. It is
expected that for too high of a doping concentration (for example, $>10^{16}$\,cm$^{-3}$),
 the quantum efficiency
may decrease due to reduced carrier collection arising from shorter diffusion lengths, and may also
induce more SRH recombination within the junction due to doping induced defects. Both of these would
have a strong impact on the noise equivalent irradiance of such a device. As for the forward current, it
fits the ideal diode current equation with an ideality factor of 1. In other words, the forward current is
dominated by the diffusion originating in the neutral InGaAs layer based on the lifetime of holes and the
volume of the InGaAs layer.

\section{Conclusions}

    A simple model is proposed to extract minority carrier diffusion
 lengths in low doped ($<10^{16}$\,cm$^{-3}$) n-InGaAs using simple
 device measurements of long and thin diffused junction diodes separated by various distances.
The reported diffusion lengths range between $70\,-\,85$\,\textmu m for an
 InGaAs doping concentration between $5\,-\,9\times10^{15}$\,cm$^{-3}$.
 The ensuing analysis also yields minority carrier mobilities which
 range between $500\,-\,750\,$ cm$^2$/V$\cdot$s, again depending on the doping
 concentration of the InGaAs. The proposed methodology can
easily be applied to any semiconductor device with clearly defined diffused junctions, although the interdiode distances must be chosen
carefully such that the maximum interdiode distance is on the order of 3 diffusion lengths. The
extracted diffusion lengths represent an effective value which is dominated by heavy holes, since the
heavy holes have a much larger density of states than light holes. The correlation between these
diffusion lengths and dark current characteristics of 100 pixel test arrays is also given, whereby both
doping concentrations result in very similar dark current densities due to counter-balancing contributions from junction
generation-recombination via SRH and diffusion in the neutral region. Overall, nearly half of the dark current
can be attributed to junction generation-recombination, whereas the remaining half is due to diffusion
processes in the neutral InGaAs region.

\acknowledgments 

The authors would like to thank Craig Storey for useful discussions.


\end{document}